\documentclass[10pt,aps,prb,final,twocolumn,amsmath,amssymb,superscriptaddress,longbibliography]{revtex4-1}
\usepackage[pdftex, colorlinks=true, linkcolor=blue, citecolor=blue, urlcolor=blue]{hyperref}
\usepackage{graphicx}
\usepackage{aurical}
\usepackage[T1]{fontenc}
\usepackage{units}
\usepackage{siunitx}
\usepackage[version=3]{mhchem}

\graphicspath{{./figures/}}
\DeclareGraphicsExtensions{.png,.pdf}

\newcommand{\kaust}{King Abdullah University of Science and Technology (KAUST)\\
	Physical Science and Engineering Division (PSE),
	Thuwal, 23955-6900, Saudi Arabia}
\begin{document}
	
	\title{Noncollinear antiferromagnetic textures driven  high-harmonic generation from magnetic dynamics in the absence of spin-orbit coupling}
	\author{Ousmane Ly}
	\email{ousmane.ly@kaust.edu.sa}
	\affiliation{\kaust}
	
\begin{abstract}
We demonstrate the generation of high order harmonics in carrier pumping from precessing ferromagnetic or antiferromagnetic orders, excited via magnetic resonance, in the presence of topological antiferromagnetic textures. This results in  an enhancement of the carrier dynamics by orders of magnitude, enabling for an emission deep in the THz frequency range. Interestingly, the generation process occurs in an intrinsic manner, and is
solely governed by the interplay between the s-d exchange coupling underlying the noncollinear antiferromagnetic order and the dynamical s-d exchange constant of the magnetic drive. Therefore,  the relativistic spin-orbit interaction is not required for the emergence of  high harmonics in the pumped currents. Accordingly, the noncollinear topological antiferromagnetic order is presented as an alternative to spin-orbit interaction for the purpose of harnessing high harmonic emission in carrier pumping. Furthermore, we demonstrate the emergence of high harmonics from random magnetic defects. This suggests the universality of the magnetically induced high harmonic emission in the presence of real and/or momentum space noncollinear textures.
Our proposal initiates a tantalizing prospect for the utilization of topological textures in the context of the highly active domains of ultrafast spintronics and THz emission.
\end{abstract}
	
\maketitle

\section{Introduction }
High harmonic generation (HHG) is a mechanism by which a frequency can be converted into its higher multiples, usually under optical excitations. In the field of quantum optics, HHG has been proposed to convert laser light from infrared or ultraviolet range to the soft X-ray window \cite{Popmintchev2010, Popmintchev2012}. This has led to unprecedented spectroscopic tools allowing to probe electron dynamics at the attosecond time scale \cite{Li2020, Ghimire2022}. 
While in the attosecond physics a wavelength enhancement of few orders of magnitude can be obtained via HHG \cite{Popmintchev2010, Popmintchev2012},
in solid state systems the generated high harmonics are limited in both number and intensities \cite{Ghimire2019, Jongkyoon2022}.

In a very recently proposed approach \cite{Ly2022}, the relativistic spin-orbit interaction has been combined with magnetic resonance induced precession to trigger ultrafast dynamics in both charge and  spin currents. Interestingly, the use of spin pumping as a driving force instead of optical excitations opens up tantalizing routes toward engineering ultrafast spintronic devices with the grand advantage of being highly scalable. 
This is in contrast to opto-magnetic studies \cite{Beaurepaire1996, Kampfrath2013, Kudasov2018, Quessab2019, Deb2019, Deb2019b}, where a laser beam is used to excite ultrafast carrier or magnetic dynamics.
For instance, in the recent proposal by  Kampfrath {\it{et at}} \cite{Kampfrath2013},  a femtosecond laser beam is used to generate non-equilibrium electron distributions in a geometry constituted of a thin ferromagnetic iron film combined with a thin noble metal layer. 
The spin currents generated in these magnetic heterostructures, and measured via inverse spin Hall effect, have been found to display THz dynamics with a wide broadband spectrum of about 30 THz. Introducing the relativistic spin-orbit interaction, our approach has allowed for a much wider spectrum. Upon tuning the spin-orbit coupling strength, an enhancement of the bandwidth by an order of magnitude has been enabled.

Although the Rashba induced ultra-high frequency generation, provides an interesting alternative to the laser driven ultrafast dynamics, the necessity of fine tuning the underlying spin-orbit coupling strength to excite the strongly non-linear regime would pose a grand challenge from the experimental standpoint.

The key ingredient required for the emergence of HHG from magnetic dynamics is the presence of an interaction that leads to 
spin-flip scattering events within adiabatic magnetic cycles, a scenario guaranteed by the relativistic spin-orbit interaction for instance.
Accordingly,  
topological antiferromagnetic textures \cite{Smejkal2018, Bonbien2021} appear to be natural alternatives to the spin-orbit coupling systems in this perspective. As a matter of fact, their real space noncollinear magnetic arrangement would trigger spin-flip scattering events.  


In the present article, we propose to exploit coplanar and non-coplanar  antiferromagnetic textures to induce very high harmonics in carrier dynamics resulting from spin pumping out of a precessing magnetic order, excited via ferromagnetic or antiferromagnetic resonance.
Whereas in conventional spin pumping \cite{Tserkovnyak2002, Cheng2014},  a rotating magnetic moment pumps a spin current whose dynamics operate at the frequency of the underlying magnetic resonance, in the presence of noncollinear antiferromagnetic order, the spin rotation symmetry is broken. Therefore, the spin is not conserved, and spin-flip scattering events occur in a fashion similar to the case of the relativistic spin-orbit interaction. This similarity between the noncollinear order and spin-orbit coupling has been used to justify the emergence of unconventional spin Hall currents in spin-orbit coupling free antiferromagnets \cite{Zhang2018}. Further, the scattering events resulting from the topological order are accompanied by the emission of high frequencies, multiples of the driving frequency.
Accordingly, an enhancement of the carrier dynamics by orders of magnitude is predicted. Hence, the resulting currents are emitted at frequencies far beyond the intrinsic THz dynamics that has made the antiferromagnetic materials very popular in the field of spintronics.

As a proof of principal of the HHG effect in noncollinear antiferromagnets, we present numerical simulations of the charge emission in this class of magnetic materials with a particular focus on the role of the s-d exchange parameter and the driving frequency in controlling the emission's bandwidth. 

The manuscript is organized as follows. First, the numerical implementation of the systems of interest is described in Sec. \ref{sec:methodo}. The numerical results are subsequently presented in Sec. \ref{sec:results}. Further, a discussion of our findings is provided in Sec. \ref{sec:discussion}, followed by conclusions and perspectives given in Sec. \ref{sec:conclusion}.

\section{Methodology} 
\label{sec:methodo}
We consider the following tight binding Hamiltonian
\begin{equation}
\label{eq:ham0}
H_0=\Delta\sum_{i}  c_i^{\dag} (\mathbf{S}_i \cdot \mathbf{\sigma} )c_i -\gamma \sum_{<i,j>}c_i^{\dag} c_j,
\end{equation} 
where $c_i^{\dag}$ and $c_i$ are respectively the creation and annihilation operators at the site $i$ and $\mathbf{S}_i$ stands for the magnetic moment at the same site. The strength of the coupling between the magnetic texture and the localized electrons is given by the constant $\Delta$ expressed in units of the hopping energy $\gamma$. Here, $\mathbf{\sigma}$ is the vector of Pauli matrices.
In ref \onlinecite{Ndiaye2019} a unitary transformation has been used to eliminate the off diagonal terms at the magnetic sites. A transformation of the form $U=n_i\cdot\mathbf{S}_i$ has been introduced with $n_i=\mathbf{z}+\mathbf{S}_i/|\mathbf{z}+\mathbf{S}_i|$.
This transforms the  Hamiltonian to

\begin{equation}
\label{eq:htilde}
\tilde{H}_0=\Delta\sum_{i}  c_i^{\dag} \mathbf{\sigma}_z c_i -\gamma \sum_{<i,j>}c_i^{\dag} (n_i.\sigma)(n_j.\sigma)c_j.
\end{equation} 
In general, replacing $z$ by $x$ or $y$ directions would transform the Hamiltonian into a form similar to Eq. \eqref{eq:htilde} with modifying $\sigma_z$ by $\sigma_x$ or $\sigma_y$ respectively.
Therefore, the system can be seen as an effective ferromagnet with a spin orbit like coupling expressing the interaction of spin with the lattice sites (second term of Eq. \eqref{eq:htilde}). 
A similar transformation has been introduced to provide an explanation of the anomalous Hall effect in the coplanar kagome antiferromagnetic system \cite{Busch2020}.  

Depending on the type of the considered noncollinear texture, the magnetic moments in Eq \eqref{eq:ham0} are  implemented accordingly.
To account for the time dependent magnetic precession, we consider a magnetization vector in the form 
\begin{equation}
	\label{eq:mt}
	\mathbf{m}(t)=(\sin{\theta}\cos{\omega }t, \sin{\theta}\sin{\omega t}, \cos{\theta}), 
\end{equation} 

where $\omega$ and $\theta$ stand respectively for the precession frequency and the cone opening of the driving dynamics. The corresponding dynamical s-d exchange term reads
 \begin{equation}
 \label{eq:vt}
 {V}(t)=J\sum_{i}  c_i^{\dag} (\mathbf{m}(t) \cdot \mathbf{\sigma} )c_i, 
 \end{equation} 
 with $J$ being the s-d exchange energy of the dynamical ferromagnetic order.
 Therefore, the total time dependent Hamiltonian is given as
 \begin{equation}
 \label{eq:total}
 \mathcal{H}(t)  = H_0 + {V}(t).
 \end{equation} 

  
To compute the non-equilibrium current responses, the stationary scattering modes of the tight binding system at different energies are obtained using kwant \cite{kwant}. Subsequently, they are evolved forward in time by solving the time dependent Schr{\"o}dinger equation \cite{Kloss2021}. Hence, the time dependent scattering wave-function at site $i$ of a given mode $m$ impinging from lead $l$ with a transport energy $\varepsilon$, denoted by $\Psi_{l m\varepsilon}^{i}(t)$, can be obtained.
Accordingly, the zero temperature time dependent current flowing from between two sites $i$ and $j$ is computed as 

\begin{equation}
	\label{eq:it}
	{ {I_{ij}} }(t) = \sum_{l m} \int \frac{d \varepsilon}{\pi} \Im \{\Psi_{l m\varepsilon}^{i \dagger}(t) \mathcal{H}_{ij}(t)\Psi_{l m\varepsilon}^{j}(t)\},
\end{equation}
where $\mathcal{H}_{ij}$ is the Hamiltonian matrix element between sites $i$ and $j$ and $\Im$ stands for the imaginary part.
The total current $I(t)$ flowing from the scattering region to a given lead is obtained by  summing the current \eqref{eq:it} along the corresponding interface.
A crucial task of the numerical calculations consists of evaluating the energy integrals involved in Eq. \eqref{eq:it}.  To do so, the Gauss-Kronrod quadrature is used to  integrate from the bottom of the bands to the underlying Fermi energy. This is repeated for each transport mode of the leads. To ensure convergence, the initial interval is subsequently subdivided until the quantity of interest is stabilized or till a given tolerance is reached. 
Further, to obtain the charge currents in the frequency domain, a discrete fast Fourier transform of the time dependent signal is performed.
In the data presented throughout the text ${I}_{\omega}$ represents the absolute value of the normalized Fourier transform of ${I}(t)$. 

\begin{figure}
	\includegraphics[width=0.5\textwidth]{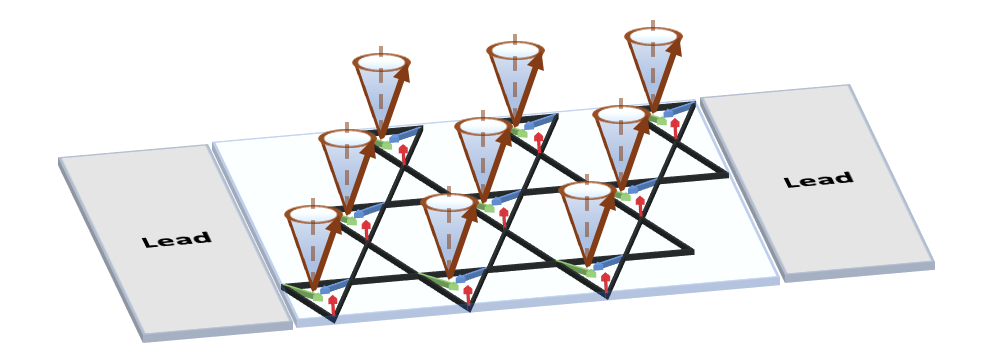}
	\caption{A sketch of the system under consideration. A noncollinear antiferromagnetic texture is exposed to the precessing ferromagnetic order (brown cones). The system is attached to two normal leads (gray area). As an illustration, a coplanar noncollinear order in the $120^\circ$ configuration is shown. The magnetic unit cell contains three magnetic moments, where each sub-lattice site hosts a precessing magnetic order (brown cone).}
	\label{fig:sketch}
\end{figure}

\begin{figure}[h!]
	\includegraphics[width=0.5\textwidth]{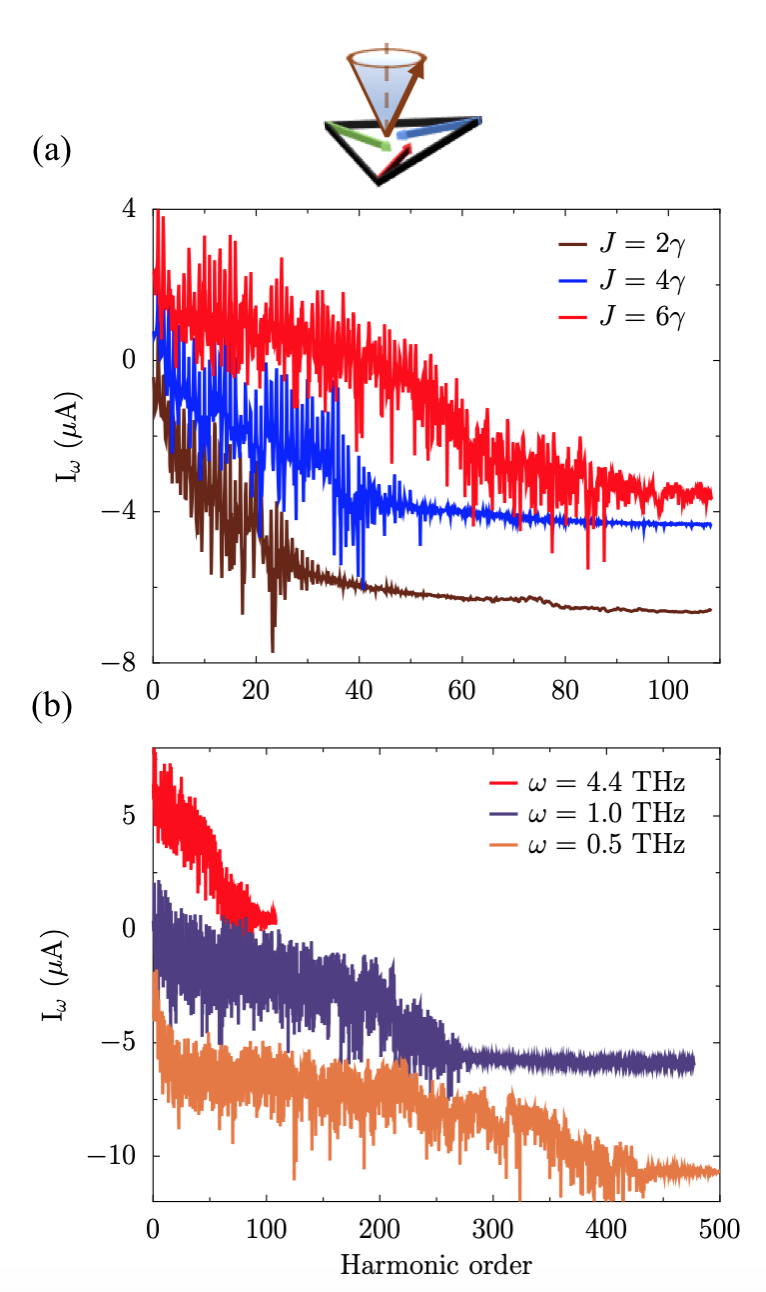}
	\caption{The Fourier transform of the charge current emitted out of a precessing magnetic order in the presence of a coplanar antiferromagnetic texture is presented in logarithmic scale. Here, a kagome lattice is considered with the magnetic moments making $120^\circ$ with respect to each other. The s-d exchange energy of the antiferromagnet is taken as $\Delta = 3 \gamma$. In panel (a), the frequency domain charge current at different ferromagnetic exchange energies are shown. The frequency of the exciting dynamics is $\omega=4.4$ THz. In the lower panel the same quantity is shown at different driving frequencies. The dynamical ferromagnetic exchange energy is taken as $J=6\gamma$. A precession angle of $10^\circ$ is considered in the data underlying both panels. For clarity of presentation, the signals are vertically shifted.}
	\label{fig:kagome}
\end{figure}

\section{Results }
\label{sec:results}
\subsection{High harmonic generation from ordered noncollinear antiferromagnetic textures}
\label{subsec:ordered}
We start by considering  a coplanar  kagome antiferromagnetic  system, where each couple of classical magnetic moments in a unit cell form an angle of $120^\circ$.
A sketch of the corresponding system is shown  in Fig. \ref{fig:sketch}, where the kagome antiferromagnet is subjected to a precessing ferromagnetic order.
The magnetic system is attached to normal leads where the pumped charge currents flow.
In the geometry considered here (see Fig. \ref{fig:sketch}), the precessing magnetic order is superposed to the noncollinear antiferromagnetic unit cells. Therefore, while the magnetization vector $\mathbf{m}$ is precessing in time with a frequency $\omega$, a spin current is pumped within the antiferromagnet. So far, non-equilibrium charge and spin currents flow out of the noncollinear texture according to inverse spin Hall effect.

 Fig. \ref{fig:kagome} (a) shows the high harmonic spectrum resulting from a ferromagnetic dynamics excited at a frequency of 4.4 THz. 
 In this figure different values for the s-d exchange energy of the ferromagnetic dynamics have been considered. 
 It is worth emphasizing that the high order emission is governed by three parameters. Namely, the dynamic s-d exchange coupling $J$ of the driving magnet and the s-d exchange parameter of the noncollinear antiferromagnetic system, $\Delta$, in addition to the driving frequency $\omega$. In Fig \ref{fig:kagome} (a), one can observe the enhancement of the higher order amplitudes as $J$ is increased. Nonetheless, the interplay between the dynamical exchange energy and the s-d exchange parameter of the topological texture should play a crucial role in determining the high harmonic emission process. However, the consideration of the parameter space spanned by $J$ and $\Delta$ necessitates further numerical efforts. Notwithstanding, the variation of only one parameter largely suffices as a proof of principle of the HHG effect in noncollinear antiferromagnets.  
 
\begin{figure}
	\includegraphics[width=0.5\textwidth]{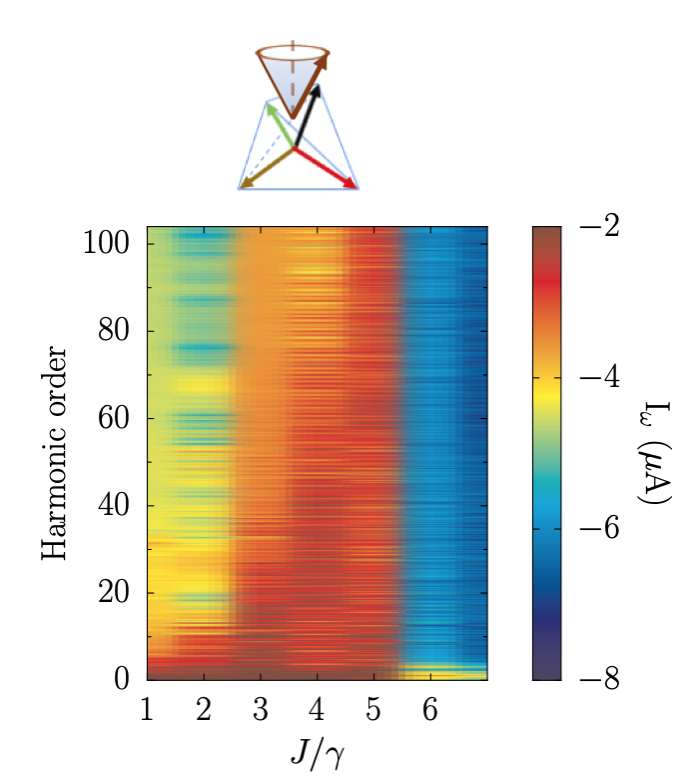}
	\caption{ The Fourier transform of the charge current emitted out of a precessing magnetic order in the presence of a non-coplanar antiferromagnetic texture is represented (in logarithmic scale) at different values of ferromagnetic s-d exchange coupling $J$. Here, an antiferromgnetic triangular  lattice in the triple-Q topological configuration is considered. The antiferromagnetic exchange coupling and the driving frequency are set at their values in Fig. \ref{fig:kagome}. A precession angle of $45^\circ$ is considered.}
	\label{fig:threeq}
\end{figure}

In Fig \ref{fig:kagome} (b), the driving frequency dependence of the charge emission in the kagome system is displayed. An increase of the high frequency cutoff at low $\omega$ values can be noticed.  
This is in consistence with theoretical results reported elsewhere in laser induced high harmonic generation \cite{Wu2015} as well as the recently reported magnetic HHG in spin-orbit coupling systems \cite{Ly2022}, where  the high frequency cutoff is found to scale as $\propto 1/\omega$. This would lead to the emission of much more harmonics at smaller frequencies. Note that the signals are vertically shifted for the clarity of presentation. Nonetheless, the evolution of the currents is not expected to be monotonous in the strongly nonlinear regime. 
Our numerical results exhibit a plateau like structure in the computed non-linear charge spectra. In ref. \onlinecite{Wu2015}, both primary and secondary plateaus were found to be featured by the harmonic emission. The emergence of these plateaus has been attributed to transitions between valance bands and specific conducting levels of the underlying Bloch electrons driven by intense laser fields.
To confirm the non-collinearity induced high harmonics, we further considered a non-coplanar antiferromagnetic texture, in the so-called 3Q topological state \cite{Shindou2001, Kurz2001, Martin2008}. In Fig \ref{fig:threeq}, the Fourier spectrum of the charge current emitted out of the 3Q texture is shown. One can clearly observe that the band width of the high frequency generation increases as a function of the s-d exchange parameter in a manner similar to the observations in the case of the kagome antiferromagnet. It is worth emphasizing, that a wide range of the parameter space is characterized by a strong high harmonic emission. However, a low harmonics response is obtained when the exchange parameter is higher than $J=5.5 \gamma$ as well as in the perturbative regime at small values of $J$. 
Notice that a large cone opening of $45^\circ$ has been considered. Although this is beyond the experimentally accessible values of the precession angle, the underlying emission illustrates the role of the cone opening in the enhancement of the high harmonics responses. 
In Fig \ref{fig:kagome} a smaller angle of precession ($10^\circ$) has been rather considered.
In order to optimize the intensities of the high order harmonics, a larger cone opening of the magnetic precession should be used. While most of experimental studies of spin-pumping are performed at relatively small angles, the possibility of achieving a cone opening that is as large as $22^\circ$ has been reported \cite{Fan2010}. 
\subsection{High harmonic emission from magnetic impurities}
\label{subsec:disordered}
In the previous subsection, ordered noncollinear antiferromagnetic systems have been considered. Bearing in mind that the effect we describe results from spin-flip scattering events, we expect that HHG should be obtained in the presence of magnetic impurities. To demonstrate this, we  consider a magnetic system, where three-dimensional magnetic moments are placed on a two-dimensional tight-binding square lattice. 

We consider classical vectors whose components are randomly sampled within the interval $[-1, 1]$.
To account for the exchange coupling of these magnetic moments to the electronic background, an overall exchange coupling parameter $J$ is introduced. The magnetic system is therefore attached to two normal leads, and the non-equilibrium charge currents are obtained by evaluating the energy integral in Eq. \ref{eq:it}.

\begin{figure}
	\includegraphics[width=0.5\textwidth]{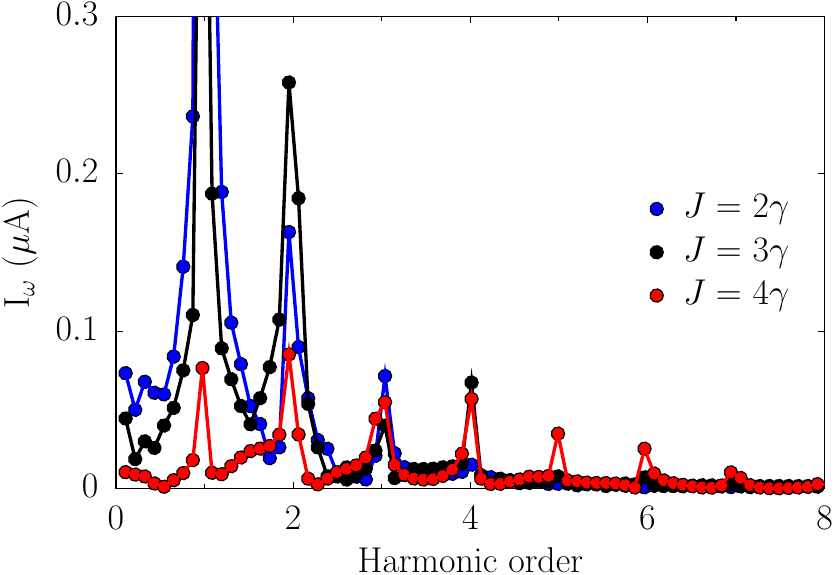}
	\caption{High harmonic spectra in the presence of magnetic disorder. Different values for the s-d exchange coupling are considered. A dynamical s-d exchange energy of $\Delta=5\gamma$ is considered. The frequency of the magnetic precession is set at $\omega = $ 4.4 THz. For clarity of presentation the first harmonic response is cut (in the case of $J=2\gamma$ and $J=3\gamma$), so that the higher harmonics are visible. Note that a linear scale is used to represent the data.}
	\label{fig:impurities}
\end{figure}

In Fig \ref{fig:impurities}, HHG spectra for different values of $J$ are displayed. Here, a high frequency of 4.4 THz is considered. 
In contrast to the ordered magnetic systems, the high harmonics spectra in the disordered texture only bear a few number of harmonics below the tenth order. Although, it is possible to enhance the emission considering lower driving frequencies, the limitation of the featured harmonics in the disordered case suggests that spin relaxation mechanisms might play an important role in the emission process. In fact, the randomization of the electronic spin in the presence of the magnetic defects would lead to the observed reduction of the high harmonic bandwidth. Consequently, the magnetically induced high harmonic generation effect constitutes a potential spectroscopic tool to study spin flip scattering events and reveal spin relaxation related phenomena in magnetic systems.

In a previous study of the magnetically induced HHG in relativistic spin-orbit coupling systems \cite{Ly2020}, it has been shown that neutral impurities do not affect the high harmonic generation spectra. Obviously, these non-magnetic defects cannot intrinsically induce high harmonic responses due to the lack of spin dependent scattering.
In the present context, we rather demonstrated the emergence of high frequency responses due to magnetic impurities in the absence of any ordered texture.  This confirms the universality of the generation mechanism in the presence of any spin-flip scattering source, although, under different circumstances and case specific parameter realization. 

It is clear that the calculations shown in Fig. \ref{fig:impurities} are in the non-perturbative regime.  Although the emergence of high order harmonics from magnetic defects is predicted, the quantitative study of the  interplay between ordered and disordered textures is beyond the scope of the actual work, which aims at presenting a proof of concept of the HHG effect in spin-orbit coupling free noncollinear magnetic systems. Nevertheless, at low impurity concentrations and at weak disorder strengths, the high frequency emission is expected to be mainly governed by the dominating topological order.

 \section{Discussion }
 \label{sec:discussion}
A particular regime in the HHG process was distinguished in the Rashba induced high order harmonics  \cite{Ly2020}. It corresponds to the case when  the spin orbit splitting is tuned to a value close to the s-d exchange energy of the driving magnet.
In such a regime, high harmonics emerge with noticeably strong amplitudes, all comparable to the fundamental frequency's response amplitude. 
 
In the case of noncollinear textures, the condition underlying the maximally excited regime is not easy to discern by numerical means. Therefore, developing an analytical theory to spot out a particular parameter range for optimizing the generation, would be of great relevance to harness the strong non-linearities in the charge responses emanating from the interaction between magnetic dynamics and noncollinear antiferromagnetic textures. The Lippmann Schwinger expansion formalism \cite{Jalabert2010, Gorini2013} would be presumably well adapted for deriving the underlying strongly non-linear spectra.  
 
 Although, the analytical description of the HHG is beyond the scope of the actual study, it is of great relevance to discuss the mechanism behind the emergence of HHG in the presence of real space noncollinearities. We shall first notice that in the absence of the noncollinear order, the time dependent spin current vector is simply given by \cite{Tserkovnyak2002} $\mathbf{m}(t)\times \partial_t \mathbf{m}(t)$, up to a DC transport pre-factor, namely, the real part of the spin mixing conductance. In the presence of a magnetic dynamics in the form given by Eq.\eqref{eq:mt}, the resulting spin current vector is obtained as 
 \begin{equation}
 \label{eq:pumping}
 \mathcal{J}\propto(\sin{2\theta}\cos{\omega t}, \sin{2\theta}\sin{\omega t}, -2\sin^2{\theta}).
 \end{equation} 
 Hence, the in-plane components of the spin current oscillate with a frequency  $\omega$ corresponding to the initial frequency of the driving dynamics.   
 Yet, in the presence  of the noncollinear order, the formula above does not apply anymore. Our results suggest that the underlying spin densities operate at a time scale much more smaller than the period of the precessing magnetic moment ($\tau =2\pi/\omega$), in the adiabatic limit. Therefore, the direction of the spin density is flipped many times within a single magnetic cycle, and each spin flip corresponds to the accumulation of a phase $\omega t$ in the electronic wavefunction. The number of times the spin density is flipped corresponds to the number of the emitted high harmonics. This mechanism of emission has been also used in the context of the spin orbit coupling induced high harmonics carrier pumping \cite{Ly2020}. It is in close analogy to the three step model usually evoked to describe the light induced HHG in gaseous media \cite{Corkum1993} and correlated electron systems \cite{Imai2020}. The phenomenological explanation above is in consistence with the emergence of high harmonics in the presence of magnetic defects (see \ref{subsec:disordered}).

Note that while our proposal is based on a pure quantum mechanical effect, recent studies have proposed the exploitation of the nonlinear character of the equations governing the classical magnetic dynamics to excite high frequencies in ferromagnetic skyrmions \cite{Rodrigues2021} or magnonic excitations \cite{Koerner2022}.
In these approaches the resulting dynamics operate at the GHz regime. Alternatively, our approach enables for an emission deep in the THz range. 
 
A key aspect that differentiates the HHG in noncollinear systems from their Rashba spin-orbit coupling counterparts is the possibility to tune the value of the main parameter which is either the antiferromagnetic s-d exchange energy or the spin-orbit coupling strength. In the latter case a potential gate can be used to adjust the Rashba energy. In contrast, the s-d exchange parameter in the former case can be adjusted by inserting an insulating barrier leading to the reduction of the s-d exchange coupling strength. 
 
Now, let us  discuss the possible experimental realization of the noncollinearity induced HHG.
A potential candidate for the observation of the proposed effect is the itinerant antiferromagnet $\gamma-$FeMn, which hosts a non-coplanar antiferromagnetic order in the triple-Q topological configuration.
 This particular antiferromagnet has been studied in the context of anomalous Hall effect resulting from the spin chirality in  the magnetic unit cells underlying its structure \cite{Shindou2001}. Additionally, the class of Mn compounds including Mn$_3$Sn \cite{Nakatsuji2015, Kimata2019} and Mn-based antiperovskite nitrides \cite{Zhou2020} provides an interesting platform for the realization of the non-collinearity induced HHG. The latter class of materials has been recently investigated in different contexts including spin caloritronics \cite{Zhou2020}, spin Hall \cite{Zhang2018, Busch2021}, anomalous Hall \cite{Chen2014} and magnetic spin Hall \cite{Zelezny2017, Kimata2019} effects.  Note that laser induced THz emission in Mn$_3$Sn has been also recently studied \cite{Zhou2019} as well as THz anomalous Hall effect \cite{Matsuda2020}.
 So far, time resolved inverse spin Hall effect has been widely investigated in broadband THz emission measurements \cite{Kampfrath2013, Seifert2017b, Cramer2018, Gueckstock2021,  Evangelos2021}. This provides a suitable platform for probing the time dependent spin-charge conversion signals, which should feature the strongly non-linear carrier dynamics. Furthermore, the emergence of the magnetic spin Hall effect \cite{Kimata2019} in Mn$_3$Sn provides a rather interesting perspective to observe the effect. This is particularly due to the fact that the underlying signal is odd under time reversal symmetry. Therefore, magneto-optical Kerr measurements might be used to signal the underlying ultrafast dynamics.
 
\section{Conclusions and perspectives }
\label{sec:conclusion}
In summary, we have proposed noncollinear antiferromagnetic textures as sources for the magnetically induced high harmonic generation. 
The dependence of the underlying emission on the exchange coupling parameter of the driving magnetic order has been investigated. So far, the evolution of the high frequency cutoff in terms of the initial magnetic dynamics' frequency has been presented. To support our phenomenological explanation of the high frequency emission, the emergence of HHG in the presence of magnetic impurities has been demonstrated.
While  a ferromagnetic time dependent precession has been considered in the present work, the emergence of the same set of high harmonics is expected when a collinear antiferromagnet is combined with the noncollinear texture. Particularly, the use of a ferromagnetic dynamics represents an asset, since the underlying frequency of precession would be different from the resonance frequency of the noncollinear texture. Thus the driving magnetic dynamics can be excited without altering the antiferromagnetic state.
Regarding the role of the real space topology in the generation process, a similar  emission is expected in the presence of magnetic skyrmions \cite{Litzius2021}, bimerons \cite{Jani2021} or higher order topological structures such as magnetic hopfions \cite{Sutcliffe2017, Kent2021}. 
Furthermore, we highlight that in the present work the proposed HHG is mainly governed by the noncollinear order. In contrast to a previous proposal \cite{Ly2020}, the emergence of the highly non-linear regime does not require the relativistic spin orbit interaction.
Nevertheless,  the study of the interplay between spin-orbit coupling and noncollinear antiferromagnetism in the HHG process remains an interesting theoretical task. 


The establishment of the strongly non-linear carrier pumping from noncollinear antiferromagnetic systems opens interesting prospects toward the utilization of real space topology for the development of spintronic devices operating at the ultrafast regime. Further, we anticipate that our work will certainly inspire new directions in the so-called antiferromagnetic spintronics \cite{Jungwirth2016, Jungwirth2018, Gomonay2018, Baltz2018, Smejkal2018}  and antiferromagnetic opto-spintronics \cite{Nemec2018}, where noncollinear orders continue to play a primordial role. 

\acknowledgments

This work was supported by King Abdullah University of Science and Technology (KAUST). We acknowledge computing resources on the supercomputer SHAHEEN granted by the KAUST Supercomputing Lab. We thank A. Manchon and A. Abbout for useful discussions.

\bibliography{ref_noncol} 
	
\end{document}